# Algorithm of Segment-Syllabic Synthesis in Speech Recognition Problem

Oleg N. Karpov, Olga A. Savenkova[*]

Dnepropetrovsk National University (DNU), Dnepropetrovsk, Ukraine

**Abstract.** Speech recognition based on the syllable segment is discussed in this paper. The principal search methods in space of states for the speech recognition problem by segment-syllabic parameters trajectory synthesis are investigated. Recognition as comparison the parameters trajectories in chosen speech units on the sections of the segmented speech is realized. Some experimental results are given and discussed.

## 1. Introduction

One of the results of acoustic researches for construction automatic speech recognition systems is recognition of speech on the basis of a small set of the acoustic samples corresponding to separate phonemes of the language has some difficult [Ka01, KK01, Ka05, Ko98]. Parameters of segments of a speech signal have the dependence on parameters both previous, and the successor segments, therefore to consider continuous trajectories in terms of parameters and in terms of segments more correctly. This process is represented as structural integration of the large speech units from the small speech units and the choice of the best match these speech units to group of segments of the input realization on each step of comparison. The structural integration assumes the process of integration of speech units that is proceeding until the best match will be found for all spoken message on all set of speech units of the dictionary and for all segments of the spoken message. For a word recognition are achieved the greatest reliability than phonemes recognition. The words would have such length and to be chosen in such number that it would be possible to compose any other words or phrases and sentences of them. To these requirements satisfy words-syllables that consists of two and three symbols-phonemes [Ka01]. The algorithm speech recognition as comparison the trajectories of parameters in chosen speech units on the sections of the segmented speech is proposed.

## 2. Basic Segment-syllabic algorithm overview

The segment-syllabic recognition problem is formulated according to [Ka01].
We have the dictionary $\{SL_k\}$ that consist of $N$ syllables. The reference sequence of parameters (the trajectory of parameters [KS04]) $Y_k = (y_{k1}, y_{k2}, \ldots, y_{km_k})$ is set for each syllable $SL_k$, where $m_k$ is length (number of points) of the trajectory of parameters for the $k$-th syllable. Each syllable $SL_k$ contains $n_k$ symbols-phonemes $\alpha_j^k$ ($k = 1 \div N$, $j = 1 \div n_k$). Each trajectory of parameters $Y_k$ contains $m_k$ elements which is united in $n_k$ segments $SG_j^{y_k}$ for corresponding symbols-phonemes $\alpha_j^k$.

---

[*] e-mail: 2sol@ukr.net

On the input we have a sequence $X = (x_1, x_2, \ldots, x_r)$. The input sequence $X$ is segmented on $p$ segments-phonemes $SG_l^x$ ($l = 1 \div p$). Segments-phonemes of input sequence $SG_l^x$ might be combined in M groups-syllables $X_i$ ($i = 1 \div M$). The phoneme structure of input sequence $X$ is unknown.

**Task**: to put in the best way input sequence $X$ matches to reference sequences $\{Y_k\}$, evaluating distance $d$

$$d = \sum_i \min_k (X_i \# Y_k), \qquad (2.1)$$

where $X_i$ contains segments-phonemes $SG_l^x$; $Y_k$ contains segments-phonemes $SG_j^{y_k}$; # - comparison operation (dynamic programming).

## 3. Algorithm of segment-syllabic synthesis in state space

For the best conformity of the reference speech units and the input speech signal segments the problem of segment-syllabic synthesis for each speech unit and for all trajectory as a whole is solved. The problem of segment-syllabic synthesis consists of searching a such reference trajectory of parameters $X^*$ for which the best match with an input speech signal trajectory of parameters $X$ on all set of syllables is get.

The concatenated syllables-patterns $SL_k^*$ trajectories of parameters $Y_k^*$ are chosen in accordance with the following algorithm.

For each input speech signal syllable $X_i$ ($i = 1 \div M$) and a reference syllable $SL_k$ is required:

$$d_i = \min_k (X_i \# Y_k) = \min_k \sum_{l,j} \left( SG_l^x \# SG_j^{y_k} \right). \qquad (3.2)$$

The best approximation of the realization $X$ on all set of syllables is minimization of value

$$d = \sum_{i=1}^{M} d_i. \qquad (3.3)$$

This problem definition of segment-syllabic recognition to find components $Y_k^*$ for a reference trajectory $X^*$ concatenation is required to carry out exhaustive search variants-combinations of trajectories $\tilde{X}^*$ and calculation of corresponding distances. It needs huge time expenses.

Reduction set of considered variants $\tilde{X}^*$ for the input speech signal, worst case time and worst case space can be achieved due to use the effective artificial intellect algorithms. For this purpose we formulate a segment-syllabic synthesis problem as state space search [Br04, RN06].

As a state $M$ in a segment-syllabic synthesis problem (the state characterizes some problem solution) we mean a combination of syllables-patterns $\{Y_k\}$ which are contained in the synthesized trajectory of parameters for the input speech signal

$$\tilde{X}_M^* = (Y_1, Y_2, \ldots, Y_i, \ldots, Y_M).$$

Initial state: no syllable-pattern has been matched. Nodes (vertex) numbering of the synthesis graph is correspond to the segments-phonemes numbers of the input speech signal $SG_l^x$ ($l = 1 \div p$). Each arc $l_i \to l_{i+n_k}$ that connecting nodes of the synthesis graph (the possible transitions between states) has an associated weight (costs) $d_{i,\,i+n_k}$. The weight $d_{i,\,i+n_k}$ is the minimal distance value between syllable $X_i$ of input realization and the reference syllable $Y_k$ ($k = 1 \div N$) calculated according to (3.2). The integrated similarity of trajectories $X$ and $\tilde{X}^*$ are defined the sum of costs on any path from the initial node to the final node. The target state or the problem solution is a combination $\{Y_k^*\}$, and corresponding them $\{SL_k^*\}$, on a path from an initial segment $SG_1^x$ to a final segment $SG_p^x$ of the input speech signal with the minimal value (3.3).

There are a lot of approaches to search of a solution path problem in terms of state space. The search strategy define the order of the states discovery. A selection of search strategy depends on a problem definition and the size of the state space.

Segment number of the input speech signal $SG_l^x$ ($l = 1 \div p$) define the size of the state space in a problem of speech recognition by segment-syllabic synthesis. To find a solution in the state space might be used a basic graph-searching algorithms called depth-first search (DFS) and breadth-first search (BFS). Other graph-searching algorithm is organized as elaborations of basic strategy [Br04, RN06].

A variants-combination of a trajectory of parameters $\tilde{X}^*$ is received by the procedures DFS or BFS. The trajectory $\tilde{X}^*$ consists of the syllables-patterns trajectories of parameters $Y_i$ like that

$$\tilde{X}^* = (Y_1, Y_2, \ldots, Y_i, \ldots, Y_R),$$

where $R$ is the syllables count of the trajectory $\tilde{X}^*$ (this value is correspond to the syllables number of the input trajectory of parameters).

For the given combination is required to find the reference trajectory of parameters $X^*$ on all set of syllables. To find optimum solution we introduce two models of adjustment process for the reference trajectories parameters.

# 4. Modelling of the adjustment process for the reference trajectories parameters

There is problem of parameters adjustment one trajectories of parameters to another in the best way. To find a solution of this problem we apply the mean-square approximation method with linear conditions-equations. The required smoothness the syllables-patterns trajectories of parameters in the points of merging are provided the conditions-equations. The received trajectories functions and the first derivatives of these functions should be continuous in all time interval.

Let's consider the following models of adjustment process for the reference trajectories of parameters:

- Linear model $\tilde{f}(X) = A \cdot f(X) + B$,

where $a$, $b$ are the parameters of linear model;

- Quadratic model $\tilde{f}(X) = A \cdot f^2(X) + B \cdot f(X) + C$,

where $a$, $b$, $c$ is the parameters of quadratic model, $f(X)$ is the function of the syllables-patterns trajectories of parameters before adjustment.

## 4.1 Linear model

Let's consider the linear model transformation for the syllables-patterns trajectories of parameters $Y_k$ ($k = 1 \div R$) which are combined in the synthesized trajectory of parameters $\tilde{X}^*$ like so

$$\tilde{X}^* = \begin{cases} \tilde{Y}_{1i}, & N_0 \leq i < N_1, \\ \tilde{Y}_{2i}, & N_1 \leq i < N_2, \\ \ldots \\ \tilde{Y}_{ki}, & N_{k-1} \leq i < N_k, \\ \ldots \\ \tilde{Y}_{Ri}, & N_{R-1} \leq i < N_R, \end{cases} \qquad (4.1)$$

where

$$\tilde{Y}_{ki} = a_k \cdot Y_{ki} + b_k, \; (k = 1 \div R). \qquad (4.2)$$

To find the unknown parameters of linear model $a_k$, $b_k$ ($k = 1 \div R$) used the mean-square approximation method with linear conditions-equations

$$\sigma^2(a_1, a_2, \ldots, a_R, b_1, b_2, \ldots, b_R) = \sum_{k=1}^{R} \left( \sum_{i=N_{k-1}}^{N_k} | \tilde{Y}_{ki} - Y_{ki} |^2 \right) \to min. \qquad (4.3)$$

The linear conditions-equations follow from equality of the functions values in a point of merging $T_j$ ($j = k-1$, $k = 1 \div R$)

$$\tilde{Y}_k(T_j) = \tilde{Y}_{k+1}(T_j), \qquad (4.4)$$

or

$$a_k \cdot Y_k(T_j) + b_k = a_{k+1} \cdot Y_{k+1}(T_j) + b_{k+1}. \qquad (4.5)$$

To minimize function (4.3) all partial derivatives ($R \times 2$) for $k = 1 \div R$ should be equal to zero, i.e.

$$\frac{\partial \sigma^2(a_1, a_2, ..., a_R, b_1, b_2, ..., b_R)}{\partial a_k} = 0, \qquad (4.6)$$

$$\frac{\partial \sigma^2(a_1, a_2, ..., a_R, b_1, b_2, ..., b_R)}{\partial b_k} = 0. \qquad (4.7)$$

For $k = 1 \div R$ expressions (4.6), (4.7) might be written as:

$$\frac{\partial \sigma^2(a_1, a_2, ..., a_R, b_1, b_2, ..., b_R)}{\partial a_k} = \sum_{i=N_{k-1}}^{N_k} 2 \cdot (a_k Y_{ki} + b_k - Y_{ki}) \cdot Y_{ki} = 0, \qquad (4.8)$$

$$\frac{\partial \sigma^2(a_1, a_2, ..., a_R, b_1, b_2, ..., b_R)}{\partial b_k} = \sum_{i=N_{k-1}}^{N_k} 2 \cdot (a_k Y_{ki} + b_k - Y_{ki}) \cdot 1 = 0. \qquad (4.9)$$

The equations (4.8), (4.9) are transformed to view

$$a_k \sum_{i=N_{k-1}}^{N_k} (Y_{ki})^2 + b_k \sum_{i=N_{k-1}}^{N_k} Y_{ki} = \sum_{i=N_{k-1}}^{N_k} (Y_{ki})^2, \qquad (4.10)$$

$$a_k \sum_{i=N_{k-1}}^{N_k} Y_{ki} + b_k \cdot n_k = \sum_{i=N_{k-1}}^{N_k} Y_{ki}, \qquad (4.11)$$

where

$$n_k = \sum_{i=N_{k-1}}^{N_k} 1.$$

To find unknown parameters $a_k$, $b_k$ ($k = 1 \div R$) we must solve the linear equations set ($R \times 2$):

$$a_1 \sum_{i=N_0}^{N_1} (Y_{1i})^2 + b_1 \sum_{i=N_0}^{N_1} Y_{1i} = \sum_{i=N_0}^{N_1} (Y_{1i})^2;$$

$$a_1 \sum_{i=N_0}^{N_1} Y_i + b_1 \cdot n_1 = \sum_{i=N_0}^{N_1} Y_{1i}; \qquad (4.12)$$

$$a_1 \cdot Y_1(T_0) + b_1 - a_2 \cdot Y_2(T_0) - b_2 = 0;$$
$$a_2 \cdot Y_2(T_1) + b_2 - a_3 \cdot Y_3(T_1) - b_3 = 0;$$

...

$$a_{R-1} \cdot Y_{R-1}(T_{R-2}) + b_{R-1} - a_R \cdot Y_R(T_{R-2}) - b_R = 0;$$

$$a_2 \sum_{i=N_1}^{N_2} Y_{2i} + b_2 \cdot n_2 = \sum_{i=N_1}^{N_2} Y_{2i};$$

$$a_3 \sum_{i=N_2}^{N_k} Y_{3i} + b_3 \cdot n_3 = \sum_{i=N_2}^{N_3} Y_{3i};$$

...

$$a_R \sum_{i=N_{R-1}}^{N_R} Y_{Ri} + b_R \cdot n_R = \sum_{i=N_{R-1}}^{N_R} Y_{Ri}.$$

The equations set (4.12) consist of:
the equations (4.10), (4.11) for $k = 1$;
the ($R-1$)-th equations (4.5) for $k = 1 \div R$;
the ($R-1$)-th equations (4.11) for $k = 2 \div R$.

## 4.2 Quadratic model

Let's consider the quadratic model transformation for the syllables-patterns trajectories of parameters $Y_k$ ($k = 1 \div R$) which are combined in the synthesized trajectory of parameters $\tilde{X}*$ like so

$$\tilde{X}^* = \begin{cases} \tilde{Y}_{1i}, & N_0 \le i < N_1, \\ \tilde{Y}_{2i}, & N_1 \le i < N_2, \\ \ldots \\ \tilde{Y}_{ki}, & N_{k-1} \le i < N_k, \\ \ldots \\ \tilde{Y}_{Ri}, & N_{R-1} \le i < N_R, \end{cases} \quad (4.13)$$

where

$$\tilde{Y}_{ki} = a_k \cdot Y^2{}_{ki} + b_k \cdot Y_{ki} + c_k, \ (k = 1 \div R). \quad (4.14)$$

Similarly, to find the unknown parameters of quadratic model $a_k$, $b_k$, $c_k$ ($k = 1 \div R$) used the mean-square approximation method with linear conditions-equations

$$\sigma^2(a_1, ..., a_R, b_1, ..., b_R, c_1, ..., c_R) = \sum_{k=1}^{R} \left( \sum_{i=N_{k-1}}^{N_k} \mid \tilde{Y}_{ki} - Y_{ki} \mid^2 \right) \to min. \quad (4.15)$$

The linear conditions-equations in points of merging $T_j$ ( $j = k-1$, $k = 1 \div R$ ) follow from equality of:

1) the functions values in a point of merging

$$\tilde{Y}_{k-1}(T_j) = \tilde{Y}_k(T_j), \qquad (4.16)$$

or

$$a_{k-1} \cdot Y_{k-1}^2(T_j) + b_{k-1} \cdot Y_{k-1}(T_j) + c_{k-1} = a_k \cdot Y_k^2(T_j) + b_k \cdot Y_k(T_j) + c_k; \qquad (4.17)$$

2) the derivatives of the functions of parameters in a point of merging

$$\tilde{Y}'_{k-1}(T_j) = \tilde{Y}'_k(T_j), \qquad (4.18)$$

or

$$2 \cdot a_{k-1} \cdot Y_{k-1}(T_j) + b_{k-1} = 2 \cdot a_k \cdot Y_k(T_j) + b_k. \qquad (4.19)$$

To minimize function (4.15) all partial derivatives ( $R \times 3$ ) for $k = 1 \div R$ should be equal to zero, i.e.

$$\frac{\partial \sigma^2(a_1, a_2, ..., a_R, b_1, b_2, ..., b_R, c_1, c_2, ..., c_R)}{\partial a_k} = 0, \qquad (4.20)$$

$$\frac{\partial \sigma^2(a_1, a_2, ..., a_R, b_1, b_2, ..., b_R, c_1, c_2, ..., c_R)}{\partial b_k} = 0, \qquad (4.21)$$

$$\frac{\partial \sigma^2(a_1, a_2, ..., a_R, b_1, b_2, ..., b_R, c_1, c_2, ..., c_R)}{\partial c_k} = 0. \qquad (4.22)$$

The equations (4.20), (4.21), (4.22) can be also expressed in the form as below

$$a_k \sum_{i=N_{k-1}}^{N_k} (Y_{ki})^4 + b_k \sum_{i=N_{k-1}}^{N_k} (Y_{ki})^3 + c_k \sum_{i=N_{k-1}}^{N_k} (Y_{ki})^2 = \sum_{i=N_{k-1}}^{N_k} (Y_{ki})^3, \qquad (4.23)$$

$$a_k \sum_{i=N_{k-1}}^{N_k} (Y_{ki})^3 + b_k \sum_{i=N_{k-1}}^{N_k} (Y_{ki})^2 + c_k \sum_{i=N_{k-1}}^{N_k} Y_{ki} = \sum_{i=N_{k-1}}^{N_k} (Y_{ki})^2, \qquad (4.24)$$

$$a_k \sum_{i=N_{k-1}}^{N_k} (Y_{ki})^2 + b_k \sum_{i=N_{k-1}}^{N_k} Y_{ki} + c_k \cdot n_k = \sum_{i=N_{k-1}}^{N_k} Y_{ki}, \qquad (4.25)$$

where

$$n_k = \sum_{i=N_{k-1}}^{N_k} 1.$$

To find unknown parameters $a_k$, $b_k$, $c_k$ ( $k = 1 \div R$ ) we must solve the linear equations set ( $R \times 3$ ):

$$\begin{cases}
a_1 \sum_{i=N_0}^{N_1} (Y_{1i})^4 + b_1 \sum_{i=N_0}^{N_1} (Y_{1i})^3 + c_1 \sum_{i=N_0}^{N_1} (Y_{1i})^2 = \sum_{i=N_0}^{N_1} (Y_{1i})^3 ; \\
a_1 \sum_{i=N_0}^{N_1} (Y_{1i})^3 + b_1 \sum_{i=N_0}^{N_1} (Y_{1i})^2 + c_1 \sum_{i=N_0}^{N_1} Y_{1i} = \sum_{i=N_0}^{N_1} (Y_{1i})^2 ; \\
a_1 \sum_{i=N_0}^{N_1} (Y_{1i})^2 + b_1 \sum_{i=N_0}^{N_1} Y_{1i} + c_1 \cdot n_1 = \sum_{i=N_0}^{N_1} Y_{1i} ; \\[4pt]
a_1 \cdot (Y_1(T_0))^2 + b_1 \cdot Y_1(T_0) + c_1 - a_2 \cdot (Y_2(T_0))^2 - b_2 \cdot Y_2(T_0) - c_2 = 0; \\
a_2 \cdot (Y_2(T_1))^2 + b_2 \cdot Y_2(T_1) + c_2 - a_3 \cdot (Y_3(T_1))^2 - b_3 \cdot Y_3(T_1) - c_3 = 0; \\
\quad \ldots \\
a_{R-1} \cdot (Y_{R-1}(T_{R-2}))^2 + b_{R-1} \cdot Y_{R-1}(T_{R-2}) + c_{R-1} - \\
\qquad - a_R \cdot (Y_R(T_{R-2}))^2 - b_R \cdot Y_R(T_{R-2}) - c_R = 0; \\[4pt]
2 \cdot a_1 \cdot Y_1(T_0) + b_1 - 2 \cdot a_2 \cdot Y_2(T_0) - b_2 = 0; \\
2 \cdot a_2 \cdot Y_2(T_1) + b_2 - 2 \cdot a_3 \cdot Y_3(T_1) - b_3 = 0; \\
\quad \ldots \\
2 \cdot a_{R-1} \cdot Y_{R-1}(T_{R-2}) + b_{R-1} - 2 \cdot a_R \cdot Y_R(T_{R-2}) - b_R = 0; \\[4pt]
a_2 \sum_{i=N_1}^{N_2} (Y_{2i})^2 + b_2 \sum_{i=N_1}^{N_2} Y_{2i} + c_2 \cdot n_2 = \sum_{i=N_1}^{N_2} Y_{2i} ; \\
a_3 \sum_{i=N_2}^{N_3} (Y_{3i})^2 + b_3 \sum_{i=N_2}^{N_3} Y_{3i} + c_3 \cdot n_3 = \sum_{i=N_2}^{N_3} Y_{3i} ; \\
\quad \ldots \\
a_R \sum_{i=N_{R-1}}^{N_R} (Y_{Ri})^2 + b_R \sum_{i=N_{R-1}}^{N_R} Y_{Ri} + c_R \cdot n_R = \sum_{i=N_{R-1}}^{N_R} Y_{Ri} .
\end{cases} \quad (4.26)$$

The equations set (4.26) consist of:

the equations (4.23), (4.24), (4.25) for $k=1$;
the ($R-1$)-th equations (4.17) for $k=1 \div R$;
the ($R-1$)-th equations (4.19) for $k=1 \div R$;
the ($R-1$)-th equations (4.24) for $k=2 \div R$.

## 5. Experimental research and results

The segmented trajectory of parameters X that incoming on input of the recognition system [KS04], is considered as set of 2-, 3- and 4-segmented syllables.

Let's consider the problem construction of a reference trajectory of parameters $X^*$ from some set of syllables for an input speech signal. The set of available syllables includes dictionaries $\{E_i\}$ ($i = 2, 3, 4$) which consist of 2-, 3- and 4-segmented syllables for the specified words. The words (numbers from zero to one hundred) have been analysed and as result the dictionaries of syllables are built.

A space state is presented as the synthesis graph. The structure of the synthesis graph for variants-combinations of a reference trajectory of parameters $\tilde{X}^*$ in a case of $p=7$ segments-phonemes in the input speech signal is illustrated in Fig. 5.1.

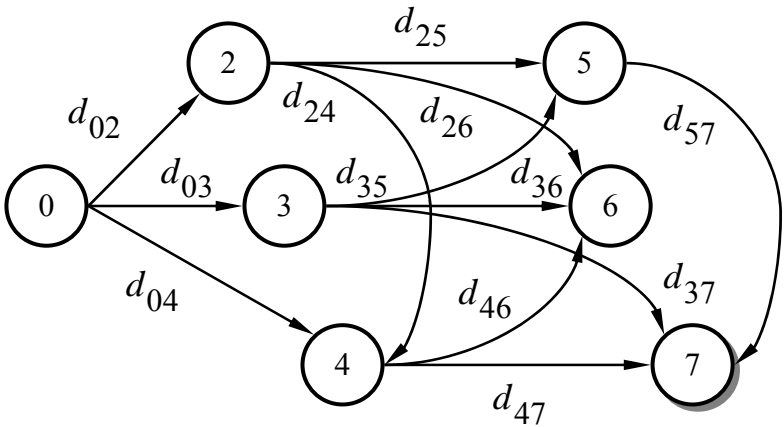

Fig. 5.1 Synthesis graph for a case $p=7$

At each level of the synthesis graph one syllable from the above mention sets of syllables in according to the chosen search strategy is added in the sequence-solution. Each arc which is starting from the node, or allowable transition, corresponds to installation of a syllable in the certain position in the sequence-solution (the position is determined by number of a segment of the input signal). The trajectory of parameters for the syllable-pattern with the minimal distance (3.2) on the each step of algorithm is used for a reference trajectories $X^*$ synthesis.

Target status: a state which is achieved on a final segment of the input speech signal with the minimal distance between the synthesized trajectory of parameters and a trajectory of the input realization on all set of syllables.

The state-space search algorithm:

(1) generate a new state, modify the existing state (change the sequence of syllables by adding a new syllable at the existing sequence-solution);

(2) check up, is the formed state the target state, and if is it so, to go to the end of algorithm else to pass to the step (1).

The search space is submitted by the following decision tree in Fig. 5.2.

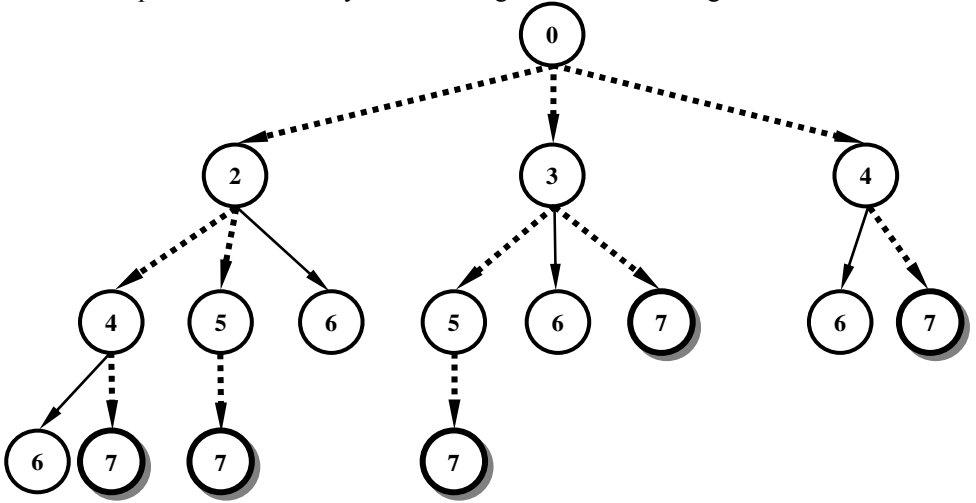

Fig. 5.2 The decision tree in a case of $p=7$ nodes of the synthesis graph

For example, for the synthesis graph in a case of $p=7$ nodes (Fig. 5.1) is required solving for the optimum for a reference trajectory of parameters in all possible variants-combinations (full search):

0-2-4-7 (2-2-3);
0-2-5-7 (2-3-2);
0-3-5-7 (3-2-2);
0-3-7 (3-4);
0-4-7 (4-3).

The **DFS** and **BFS** strategies reduce the set of considered variants-combinations. For example, for the synthesis graph of a reference trajectory of parameters (Fig. 5.1) the **DFS** and **BFS** strategies built a candidates paths which is shown in Fig. 5.3, 5.4. The results of search may depend upon the order in which the neighbors of a given node are visited.

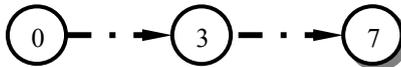

Fig. 5.3 A candidate of the solution path are produced by the BFS strategy

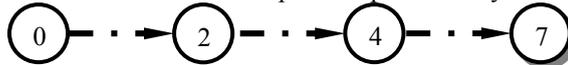

Fig. 5.4 A candidate of the solution path are produced by the DFS strategy

Researches on a finding of a reference trajectory of parameters $\tilde{X}^*$ by DFS and BFS strategies have shown:

(1) Computed value (3.3) (called solution cost), by the BFS is less than the computed solution cost by the DFS. It can be explicated that the BFS strategy guarantees to get of the shortest solution (a minimum number of transitions in state space from the initial to target), but does not guarantee to find the optimum solution.

(2) The average recognition time for the input speech signal by the segment-syllabic synthesis with strategies DFS and BFS on the order less than the average recognition time by the full search of possible combinations.

The variant-combination of a trajectory of parameters $\tilde{X}*$ is received by the procedures DFS or BFS. For the received combination to find the reference trajectory of parameters $X*$ the linear (4.1), (4.2) and the quadratic (4.13), (4.14) models of the trajectory adaptation are applied. The better results for increasing of the recognition reliability on the test set of syllables are received by the quadratic model for the trajectories adjustment.

## 5. Conclusions

The described speech recognition algorithm is the development of the algorithm proposed in [KS06A] for increasing of the recognition reliability:
(1) by using of additional criteria which estimate perspectivity of transitions between states in terms of achievement of the target state (the main difficulty of this method it is necessary to formalize the domain-dependent knowledge [KS06B]);
(2) by the adjustment process for the reference trajectories parameters.
Advantages of the adjustment process for the reference trajectories parameters in the speech recognition algorithm: the models depend on small number of linear parameters; building of the models is based on application of standard high-speed algorithms.